\documentstyle[sprocl,epsf]{article}

\def\be{\begin{equation}}
\def\ee{\end{equation}}
\def\bea{\begin{eqnarray}}
\def\eea{\end{eqnarray}}
\def\h2o{H$_2$O}
\def\rb{{{\bf r}}}

\def\xc{X$^+$(\h2o)$_n$~}
\def\yc{Y$^-$(\h2o)$_n$~}

\begin{document}

\title{{\vskip -3cm}\null\hfill 
chem-ph/9802013
\\
{\vskip 1.5cm} POSSIBLE MECHANISM OF FORMATION AND STABILITY OF
ANOMALOUS STATES OF WATER}

\author{CHEUK-YIN WONG}

\address{Oak Ridge National Laboratory, Oak Ridge, TN 37831}

\author{SHUI-YIN LO}

\address{American Technology Group, Monrovia, CA 91016\\ California
Institute of Technology, Pasadena, CA 91125}

\vspace*{0.2cm}

%%%%%%%%%%%%%%%%%%%%%%%%%%%%%%%%%%%%%%%%%%%%%%%%%%%%%%%%%%%%%%
% You may repeat \author \address as often as necessary      %
%%%%%%%%%%%%%%%%%%%%%%%%%%%%%%%%%%%%%%%%%%%%%%%%%%%%%%%%%%%%%%

\maketitle\abstracts{ We examine the physical processes which are
involved in the formation and stability of the anomalous states of
water reported recently.  The initial step of adding a small amount of
ionic compound X$^+$Y$^-$ to pure water leads to the formation of water
clusters \xc and \yc with $n>>1$.  The structure of the cluster around
the ion depends strongly on the equation of state.  We explore the
consequences of possible polymorphic states of \h2o in the liquid
phase at room temperature.  If there are low-lying polymorphic states,
the local dipole moment and the local density will change
discontinuously as a function of the radial distance from the ion, and
regions of different polymorphic states will be found at different
separations from the ion.  Fragmentation of the cluster by vigorous
shaking may break up the cluster into small domains to allow
subsequent coalescence of these domains or the growth of the domains
as seeds to form greater domains of polymorphic states.  Further
experimental and theoretical analyses are needed to study these
pictures. }

\section{Introduction} 

	Recently, experimental observations of anomalous states of
water with $I_E$ structures have been
reported~\cite{Lo96a}$^-$\cite{Lo97}.  The existence of anomalous
states of water is an interesting and new phenomenon with important
experimental and theoretical implications.  It is therefore essential
that the observations be confirmed by an independent experimental
group in order to ascertain or refute the existence of the anomalous
states.  While we await such a confirmation, it is useful to make
plausible hypotheses on the nature of the anomalous states so as to
guide further experimental and theoretical studies on this interesting
subject.

The addition of a small amount of ionic compound X$^+$Y$^-$ to pure
water is a necessary step in the production of the I$_E$ water.  We
shall first consider the formation of clusters brought about by the
addition of a small amount of X$^+$Y$^-$ in water.  In such a dilute
aqueous solution, the X$^+$ and Y$^-$ ions are well separated and
become isolated.  Stable large water clusters of the type
X$^+$(\h2o)$_n$ and Y$^-$(\h2o)$_n$ will form around the ions in the
dilute aqueous solution since similar isolated clusters of
X$^+$(\h2o)$_n$ and Y$^-$(\h2o)$_n$ are found to be stable aggregates
in many experiments~\cite{Goo70}$^-$\cite{Coe94}.  Specifically,
individual assembly of \xc and \yc, with X$^+$=H$^+$, Li$^+$, Na$^+$,
K$^+$, Ca$^{+2}$, and Y$^-$=Cl$^-$, Br$^-$, I$^-$, and OH$^-$ have
been observed~\cite{Goo70}$^-$\cite{Coe94}.
%{Goo70,Yam74,Cal70,Sch95a,Sch95b,Vig94,Yan91,Hir88,Lar84,Coe94}.
Similar cooperative effects on complexes of alcohol with proton
acceptors have also been observed~\cite{Tuc77}.  For the
H$^+$(\h2o)$_n$ cluster, a cluster size up to $n=75$ has been
identified~\cite{Sch95a}, indicating the stability of large clusters
with $n >> 1$. The limiting size of $n$ has not yet been determined.
Computer simulation shows that the six water molecules in the
Na$^+$(\h2o)$_6$ cluster prefer the 4+2 structure, with the Na$^+$ ion
in the center of the shells~\cite{Den91}.  Molecular dynamics
calculations of the hydration shell around an ion in a supercritical
aqueous solution also exhibits the clustering of \h2o molecules around
an ion~\cite{Chi97}.  One expects, therefore, that in a water solution
with a small amount of ionic substance, large \h2o clusters will form
around the ions.  These large \h2o clusters carry charges and interact
among themselves.  They may arrange themselves in an orderly manner in
the aqueous solution to give rise to clusters of even larger sizes
(superclusters) consisting of many \xc and \yc clusters.  While each
\xc or \yc may be only about 10 \AA~ in radius, it is interesting to
find out whether superclusters may be formed and may be an important
component of the I$_E$ water, where clusters of the size of 100-1000~\AA~
have been observed~\cite{Lo96a,Lo97}.  Thus, the study of these
clusters in I$_E$ water will provide information on the stability and the
interactions of the \xc and \yc in aqueous solutions.

Other interesting physics questions can also be studied with I$_E$ water.
The consideration of cluster formation leads to the examination of the
structure of the water cluster around an individual ion. One finds
that the stability of the cluster arises from the strong Coulombic
polarizing power of the ion, which attracts water molecules with their
dipole moments aligned along the electric lines of force of the ion.
The density and the average local dipole moment of the medium need not
be spatially uniform near the ion.  Their profiles depend on the
equation of state of the water medium as a function of both the
density and the average local dipole moment per molecule.

The dependence of the cluster structure on the equation of state leads
to another related and interesting subject.  It is well known that
under different conditions \h2o molecules form different stable
structures~\cite{Fle70,Mis84,Mis94} represented by different energy minima
around which the many-body state of \h2o can be locally stable.  The
study of the structure of the cluster around an ion in I$_E$ water will
provide vital information on the equation of state of water in the
dipole moment and the density degrees of freedom.  We shall see that
if there are low-lying polymorphic states in water, the dipole moment
or density will change abruptly as a function of the radial distance
from the ion.

Rigorous motion of the liquid will lead to the fragmentation of the
cluster into small domains.  The dipole moment of each fragmented
domain has been aligned by the ions.  If the equation of state of
water at room temperature has polymorphic states as exhibited by the
presence of multiple minima in densities and dipole moments, then
these small domains may coalesce to form larger domains of metastable
phases of water \cite{Lo96a}.  The domains may also be the seeds for
the growth of larger domains of the same polymorphic states
\cite{Lo96a}.  The presence of these metastable polymorphic states of
water may be important components of the I$_E$ water.  They may then
appear as an immiscible mixture without the ions in the aqueous
solution.  The existence of polymorphic states can be examined by
studying whether the medium responsible for the I$_E$ water contains the
$X^+$ and the Y$^-$ ions in the anomalous state.  This signature for
the existence of polymorphic states can be tested with future
experimental investigations.

In the following, we shall review the tetrahedron structure of a
H$_2$O molecule, the structure of a water cluster around an ion, the
existence of polymorphic states of H$_2$O in different phases, and the
possibility of multiple energy minima of the states of water with
different dipole moments and densities at room temperature.

\section{Molecular Structure of \h2o}

%\null\vskip 1.0cm
\epsfxsize=250pt \includegraphics{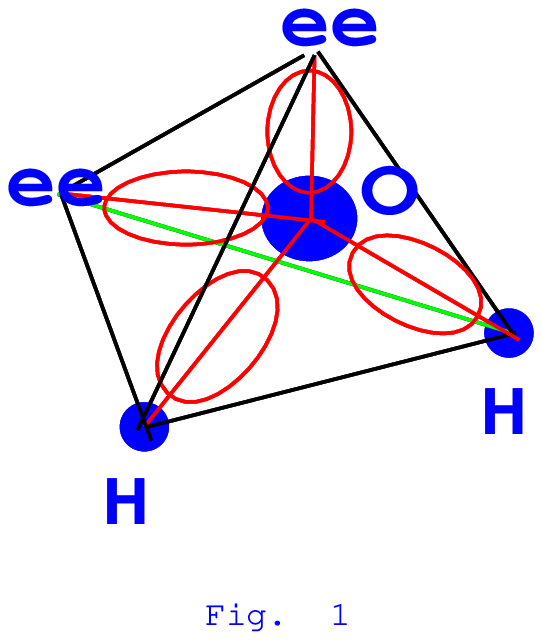}

\hangafter=-10 \hangindent=-2.2in As is well known, the H$_2$O
molecule has the structure of a tetrahedron.  The six electrons of the
oxygen atom in the outermost shell and the two electrons from the
hydrogen atoms form an octet structure of four $e e$ pairs.  The four
electron pairs coming out from the oxygen nucleus as the four supports
of the tetrahedron, with the two hydrogen nuclei at two vertices of
the tetrahedron, are shown in Figure 1.  The opening angle between the
two O-H bonds is 104.45$^{\rm o}$ as compared to $109.5^{\rm o}$ for a
regular tetrahedron.  Thus, the four vertices of the tetrahedron can
be characterized as two positively charged H vertices and two
negatively charged $ee$ pair vertices.

Because of the complex geometry of the tetrahedron, many possible
arrangements of the tetrahedron network lead to stable configurations
-- the polymorphic states.  They differ in their energies. However,
under a change of thermodynamical conditions, the lowest-lying
polymorphic state may change to lie higher than another polymorphic
state and the role of the ground state changes.  The change of
thermodynamical conditions will lead to a phase transition between
different polymorphic states.  Phase transition in which the order
parameter changes abruptly is a first-order phase transition.  Phase
transition in which the order parameter changes continuously is a
second-order phase transition.

In the normal liquid state of water, the tetrahedrons of neighboring
H$_2$O molecules are arranged in such a way that the H vertex of one
tetrahedron is in the close vicinity of the $ee$ vertex of another
tetrahedron, forming a hydrogen bond.  A hydrogen bond is weak.  It is
about 30 times weaker than a normal covalent bond, with a strength of
10-40 kJ/mol \cite{Hol96}.

	For a single \h2o molecule, there is an electric dipole moment
with a magnitude of 1.87$\times$10$^{-18}$ esu which is $0.39 |e|$\AA.
The electric dipole moment is a vector which points from the oxygen
nucleus to the mid-point between the two hydrogen nuclei.  For a
collection of \h2o molecules, one can also speak of the specific dipole
moment which can be defined as the local per-molecule average of the
vector sum of the dipole moments of a local collection of \h2o
molecules.  Because of the vector addition of the dipole moments of
individual molecules, the specific dipole moment of a many-body state
of \h2o can be quite different if the orientations of the molecules
are not random. Different arrangements of the orientation of the
molecules will lead to different specific dipole moments.  An orderly
arrangement of the molecules such that the dipole moments are aligned
nearly in the same direction will lead to an increase of the magnitude
of the specific dipole moment.

For brevity of notation, we shall use the term ``dipole moment'' as an
abbreviation to refer to the vector or the magnitude of the electric
dipole moment, the vector or the magnitude of the electric dipole
moment in units of the electron charge, or the vector or the magnitude
of the local average of the dipole moment per molecule (the specific
dipole moment), as the case may be.  The meaning of the different
cases can be simply inferred from the context.  We shall use the
unabbreviated term if its usage will help the understanding of the
discussed concepts.

\section{The Structure of a \h2o Cluster Around an Ion }

Experimental observations~\cite{Goo70}$^-$\cite{Coe94} of water
clusters of the type \xc and \yc show that the occurrence of water
molecules clustering around an ion is a common phenomenon.  This
arises from the strong Coulombic polarizing power of the ion.  For an
isolated assembly of large \xc or Y$^-$(\h2o)$_n$, is the charge
located at the center of an isolated cluster or at its surface?  The
interaction of the dipole moments of \h2o molecules with the ion
depends on the orientation of the dipole moments.  The energy of the
system is lowest when the dipole moments are aligned along the lines
of force of the ionic charge. Then the interaction between a \h2o
molecule of dipole moment $d$ with an ion of charge $q|e|$ varies with
distance to the ion $r$ as $-|q|e^2 d /r^2$, and one can show that the
energy is a minimum when the ion is located at the center.  Thus, for
a large \h2o cluster in equilibrium, the ion is located at the center.

The largest value of $n$ in the H$^+$(\h2o)$_n$ cluster which has been
observed~\cite{Sch95a} so far is 75.  What is the limiting value of
$n$ for which the X$^q$(\h2o)$_n$ cluster is still stable?  Those \h2o
molecules with a binding energy greater than the thermal energy will
not be evaporated by thermal motion.  Thus, the condition for the \h2o
at the outermost \h2o to be bound is
\begin{eqnarray}
{|q| e^2 d \over r^2 } \ge kT {~~~\rm or ~~~~} r \le \sqrt{|q| e^2 d
\over kT}.
\end{eqnarray}
For room temperature with $kT=0.025$ eV and a water molecule with
$d=0.39 |e|$\AA, the maximum radius of the cluster is $r\sim 15$ \AA~,
and the maximum number of water molecules $n$ is about 500.  If the
dipole moments are not completely aligned along the direction of the
lines of force of the ion, these values will be reduced.  One expects
therefore that the radius of the cluster is of the order of 10-15 \AA~
and a maximum cluster number of the order of 150-500.  The measured
heat of evaporation~\cite{Sch95b} of a \h2o molecule from a
H$^+$(\h2o)$_n$ cluster with $n>10$ is close to the bulk heat of
evaporation of a \h2o molecule in water in the liquid phase ($\Delta
H_v= 44.016$ kJ/mol), indicating that for large $n$, the water
molecules in the cluster are in the liquid phase.

In the water medium, the addition of a small amount of ionic compound
X$^+$Y$^-$ will lead to ions dispersed over the water medium.  When the
medium is diluted enough with a X$^+$Y$^-$ molar concentrations of
$10^{-6}$ or lower, the ions can be considered to be well separated
and isolated, and physical considerations similar to those for isolated
ions are applicable. The strong polarizing power of the ions leads to
a similar formation of \xc and Y$^-$(\h2o)$_n$.

We consider the isothermal case and place an insolated ion of charge
$q |e|$ at the origin $\rb_0=0$. To study the response of the water
medium to the presence of the charged ion, we can specify the
properties of the water medium by the ``equation of state'' function
$W(n,d)$ which is the energy per \h2o molecule.  The function $W(n,d)$
depends on the temperature, the (number) density $n$, and the specific
dipole moment $d$ (the local average dipole moment per molecule).
This function $W$ arises, at least in principle, from all interactions
between \h2o molecules.  Because of the tetrahedron structure of the
\h2o moelcule, there is the additional orientation dependence of the
equation of state on the magnitude of the average dipole moment per
molecule $d$.  The dipole moment ${\bf d}(r)$ at $\rb$, in general,
has both a magnitude $d(\rb)$ and a direction.

The total energy of the medium is
\begin{eqnarray}
\label{eq:E}
E=\int W(n(\rb), d(\rb)) n(\rb) d\rb - \int n(\rb) {qe^2 d(\rb) 
\cos \theta(\rb) \over r^2}  d\rb.
\end{eqnarray}
Note that in the above equation, the interactions between \h2o
molecules have been included in the equation of state $W(n,d)$.

The energy of the system (Eq.(\ref{eq:E})) is a minimum when the angle
$\theta$ is aligned along the electric lines of force of the ion, i.e.
$\theta=0$ for $q>0$ and $\theta=\pi$ for $q<0$.  We shall assume that
the dipole moments of the \h2o molecules have settled down such that
this alignment has been achieved.  Equilibrium is reached when the
energy is stationary upon arbitrary variation of $n$ and $d$,
$$
\delta E \!=
\!\!\!\!\int  \!\biggl [ \!{\delta \over \delta n }\biggl \{
\!W(n,d)n 
- n{|q|e^2 d(r) \over r^2 } \biggr \} \delta n
+
{\delta \over \delta d }\biggl \{  \! W(n,d)n - n{|q|e^2 d(\rb) 
\over r^2 } \biggr \} \delta d 
\biggr ] d\rb   =0.
$$
Thus, equilibrium of the medium
occurs when
\begin{eqnarray}
\label{eq:equ2}
{\delta \over \delta n }\biggl \{  W(n,d)~n - n~{|q|e^2 d(r)
\over r^2 } \biggr \}   =
\biggl \{ { \partial [W(n,d)~n]  \over \partial  n } - {|q|e^2 d(r)
\over r^2 } \biggr \}   =0
\end{eqnarray}
and 
\begin{eqnarray}
\label{eq:equ3}
{\delta \over \delta d }\biggl \{  W(n,d) - {|q|e^2 d(r)
\over r^2 } \biggr \}   =0.
\end{eqnarray}
Condition ({\ref{eq:equ2}) implies that equilibrium is reached at
those densities for which 
\begin{eqnarray}
W_{ext}  (n,d)~n = \{ W(n,d)~n - n~|q|e^2 d(r) / r^2 \}
\end{eqnarray}
is a minimum with respect to a variation in $n$, and 
({\ref{eq:equ3}) implies that equilibrium occurs for
those dipole moment values where the energy per molecule 
\begin{eqnarray}
W_{ext}  (n,d) = \{ W(n,d) - |q|e^2 d(r) / r^2 \}
\end{eqnarray}
is a minimum with respect to a variation in $d$.  The second term in
the above two equations represents the polarization force which moves
the location of the energy minima to different densities or dipole
moments.

\section{Density and Dipole Moment Profile of \h2o Cluster for a Single State} 

It is instructive to give the structure of the \h2o cluster when the
state of the water medium consists of a single state whose equation of
state is described by
\begin{eqnarray}
W(n,d) = {1\over 2} K_n {(n-n_0)^2 \over n} +{1\over 2} K_d (d-d_0)^2
,
\end{eqnarray}
where $n_0$ is the equilibrium density, $d_0$ is the equilibrium
specific dipole moment, $K_n$ is stiffness parameter with respect to
the density variation and is related to the compressibility of water,
and $K_d$ is the stiffness parameter with respect to the dipole moment
variation.  We ssume again the alignment of the dipole moments around
the ion, and we look for solutions of the density and the dipole moment
around the equilibrium values.  Then, Eq. (\ref{eq:equ2}) gives the
density profile of the cluster as
\begin{eqnarray}
n(r)=n_0+ {\sqrt{|q|e^2 d(r) / K_n} \over   r}.
\end{eqnarray}
Thus, the deviation of the cluster density from the equilibrium
density is inversely proportional to the radial distance from the ion,
depending on the compressibility $K_n$ of water.  Equation
(\ref{eq:equ3}) leads to the dipole moment profile of the cluster as
\begin{eqnarray}
d(r)=d_0+ {|q|e^2 \over  K_d r^2}.
\end{eqnarray}
The deviation of the cluster dipole moment from equilibrium is
inversely proportional to the square of the radial distance from the
ion, depending on the stiffness of the variation of $W$ with respect
to $d$.

From this analysis, one notes that the density and the magnitude of
the dipole moment of the \h2o cluster are greater than the
corresponding quantities in the surrounding medium.  It will be of
interest to see whether such a difference may lead to observable
sinking of the clusters due to gravity.

It is of interest to note the differences of two stiffness parameters
$K_n$ and $K_d$.  The stiffness parameter $K_n$ involves the
compression of the medium.  To compress water in the liquid phase,
considerable energy is needed to overcome the repulsive overlap of the
electronic densities of the molecules.  One expects that the water
medium is quite stiff against a change in density.  On the other hand,
the change of the specific dipole moment arises from the change of the
local average of the vector sum of the dipole moments of the \h2o
molecules, which can be brought about by reorienting the \h2o
molecules without moving their centers-of-mass.  As the energy
involved in making a rotational motion is considerably smaller than
the energy required to bring two molecules closer than the equilibrium
separation, one expects that the water medium is much softer against
dipole moment distortions as compared to density distortions.  That
is, it is easier to change the specific dipole moment than the
density.

\section{Polymorphic States of \h2o in the Liquid Phase}

The dipole moments for normal water are randomly oriented, and the
(average) specific dipole moment is essentially zero.  Other
arrangements of the \h2o moelcues different from that of the norm
state of water are possible, and they may lead to polymorphic states of
\h2o with different configurations, densities and dipole moments.
Polymorphic states of \h2o in the solid phase manifest themselves in
different structures of ice crystals where the tetrahedron structure
of the molecule is maintained by joining the oxygen nuclei in
tetrahedron network patterns, with a hydrogen nucleus between each
link of two oxygen nuclei.  These polymorphic states are well
known~\cite{Fle70}.  We are naturally more interested in polymorphic
states of \h2o in the liquid phase.

Polymorphic states of \h2o in the liquid phase are found
experimentally in supercooled water by Mishima
$et~al.$~\cite{Mis84,Mis94} and shown theoretically by Poole
$et~al.$~\cite{Poo97,Sci97}, and by Roberts $et~al.$~\cite{Rob97}.
The polymorphic states of supercooled water exist as the low-density
amorp (LDA) and the high-density amorp (HDA).  The transition between
the two polymorphic states has been observed to occur reversibly and
abruptly at about 135$^{\rm o}$ K and about 0.2 GPa with a volume
change of 0.02 cm$^3$/g and some hysteresis~\cite{Mis94}.  The LDA has
a volume-pressure relation very similar to Ice Ih, while the HDA is
similar to Ice V and VI at supercooled temperatures (see Fig. 1 of
Ref. [~\cite{Mis94}~]). One can interpret LDA and HDA as
configurations of excited energy minima in the liquid phase in which
the normal ground states at that temperature are different
configurations of ice in the solid phase.  For a given temperature in
which LDA (or HDA) is the lower energy ground state, the transition
from the other excited HDA state (or LDA) will involve the release of
heat, and such a release of heat has been observed
experimentally~\cite{Mis94}.

\vskip -0.00cm It is worth noting that metastable states in different
energy minima have been found in many systems in physics.  Hill and
Wheeler discussed unstable shape isomers which differ from the stable
ground state by their quadrupole moment~\cite{Hil53}.  In this case,
the quadrupole moment plays the role analogous to the order parameter
in thermodynamic systems.  Isomers with different quadrupole moments
have been found experimentally in many nuclei and can be understood as
the manifestation of multiple minima in the energy surface as a
function of the quadrupole moment~\cite{Bra72}.  They are locally
stable because they reside in the energy minima associated with their
own local variations.

\vskip -0.00cm Recent theoretical work on supercooled water indicates
that there are polymorphic states of water in which the local density
can be an order parameter~\cite{Poo97,Sci97,Rob97}.  The molecular
dynamics computer simulations of Sciortino $et~al.$~\cite{Sci97} show a
first-order liquid-liquid phase transition for water molecules
interacting among themselves with the ST2 model interaction.  These
two phases can coexist at different spatial locations at a range of
temperatures.  They differ by about 15\% in density and have different
local structures, local dynamics, and mobility, with the molecules in
the high-density phase much more mobile than the molecules in the
low-density phase.  In separate investigations on a supercooled,
dense, equilibrium Lennard-Jones liquid using molecular dynamics, it
is found that the more mobile particles form large-scale quasi-stable
string-like clusters and their dynamics is correlated in a string-like
motion~\cite{Kob97,Don97}.  The fraction of these large clusters is
about 5\% of the liquid in the model considered.

Recently Roberts $et~al.$ found from theory and simulations of
network-forming liquid that the polymorphic states occur quite
generally in systems in which the molecules interact via strong
directional intermolecular forces, as in the case of the tetrahedron
\h2o molecules or Si atoms~\cite{Rob97}.  The qualitative
characteristics of the phase transition depend on the details of
bonding and on the choice of the model parameters.  For example, a
calculated temperature-density phase diagram of a model network can
exhibit a liquid-liquid phase transition in a pure substance, in
addition to a vapor-liquid phase transition.

%\null\vskip 1.0cm
\epsfxsize=250pt \includegraphics{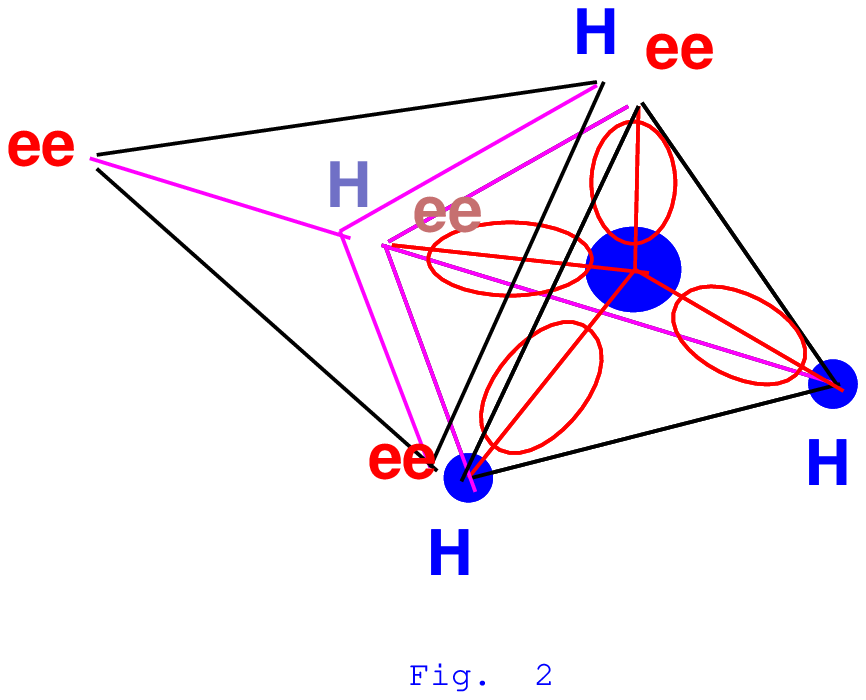} 

\hangafter=-12 \hangindent=-2.7in It will be of interest to study many
tetrahedron networks similar to those of the ice crystals to see
whether they lead to polymorphic states in the liquid phase.
Furthermore, in view of the elongated shape of the clusters as
observed in I$_E$ water samples~\cite{Lo96a,Lo97}, it will be useful
to study a network in which the simplest unit consists of two \h2o
molecules with the H vertices of these molecules in the vicinity of
the ee vertices of the other molecule.  The dipole moment of the whole
complex is aligned along the symmetry axis of the two molecules, as in
Figure 2.  The repeated and parallel arrangement of these basic units
along the same direction may lead to a state with a preferred
direction as the symmetry direction, and may exhibit an elongated
string-like behavior.

\section{Structure of \h2o Cluster for Polymorphic States}

We note in the last section the experimental observations of the
existence of the polymorphic states in supercooled
water~\cite{Mis84,Mis94}.  Theoretical supports for the occurrence of
these polymorphic states in water comes from the directional nature of
the interaction of tetrahedron molecules~\cite{Poo97,Sci97,Rob97}.
There is also the occurrence of large-scale quasi-stable, string-like
clusters in a simple Lennard-Jones liquid~\cite{Kob97,Don97}.  It is
therefore useful to explore the possibility of polymorphism for water
at room temperature.  The strong polarizing power of the ion makes it
a useful tool for such an investigation.

\hangafter=0 \hangindent=-2.7in Accordingly, we consider the
possibility of polymorphic states of water at room temperature by
representing the equation of state of water with an energy surface
containing low-lying secondary\break minima where the normal ground state
has order parameter $d_0$, and metastable states with order parameter
$d_1$ and $d_2,...$.  The secondary minima are the anomalous states
differing from the normal ground state water configuration by their
order parameters which can be either the density or the specific
dipole moment.  Each energy minimum represents a configuration with
distinct characteristics of the medium.  The case for the dipole
degree of freedom is shown schematically as the solid curve in Figure
3.

%\vskip 6cm
\epsfxsize=400pt
\includegraphics{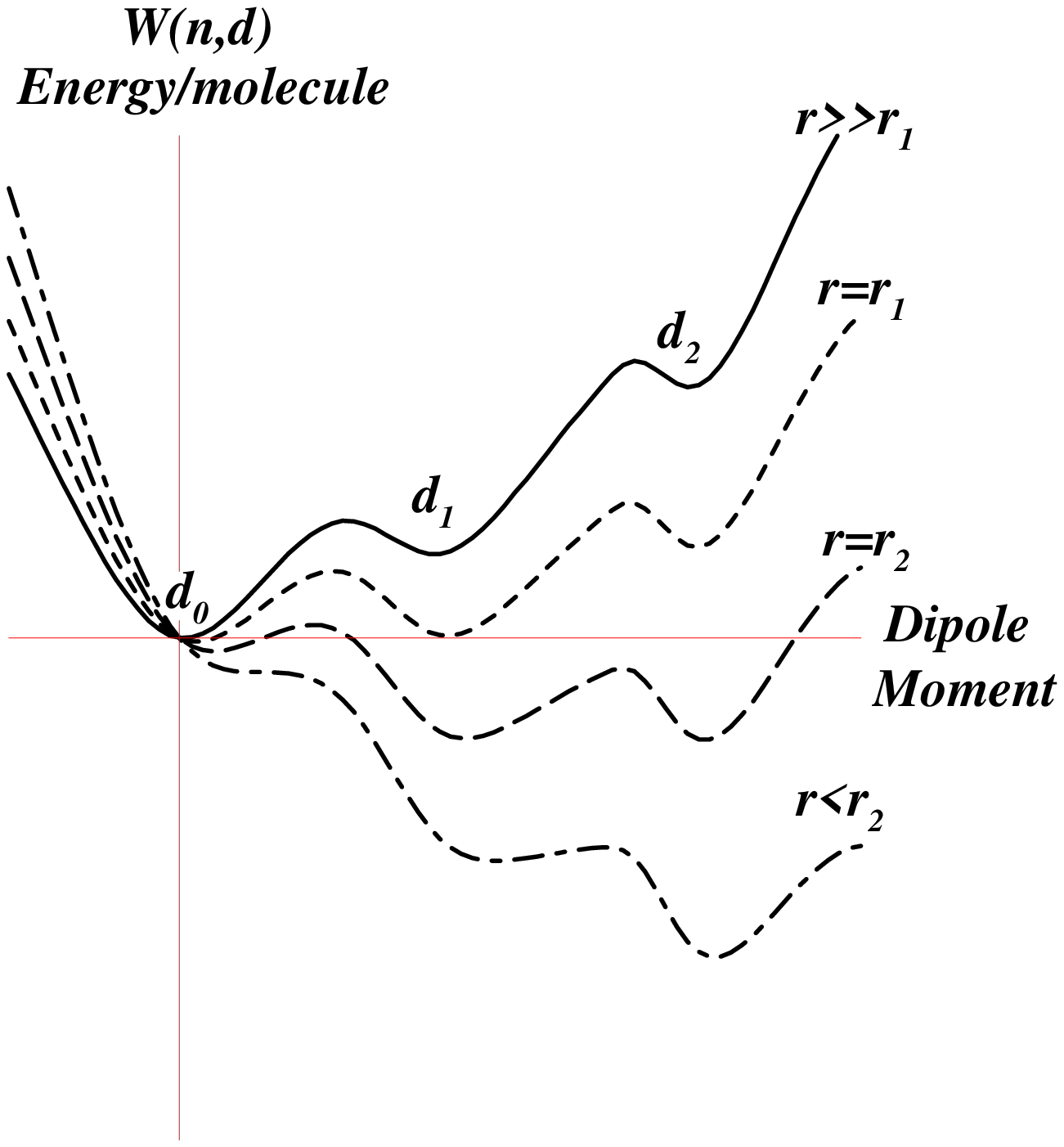}
\vskip -1.8cm \hskip 5.6cm
\begin{minipage}[t]{2.in}
{\small
\noindent  Fig.3. 
{ Energy per molecule as a function of the dipole moment
for different distances $r$ from the ion.  }
}
\end{minipage}

\vskip 0.7cm 

If the rate of change of the order parameter is rapid, the water
molecule system will jump from the lowest energy surface to the higher
energy surface, retaining the characteristics of its initial
configuration, as in the case of large Landau-Zener jump
probabilities~\cite{Lan32}.  On the other hand, if the rate of change
of the order parameter is small, then the intrinsic energy surface of
the water molecule will follow the lowest surface represented by the
solid curve in Figure 3, and the configuration of the system may
change slowly from one minimum to the next as the dynamics evolves.
In the polarization stage of the process, we shall consider only this
case of slow change of the order parameters.  The anomalous states
represented by the secondary minima at $d_1$ and $d_2$ are separated
from the minimum for the normal state at $d_0$ by a barrier.

Assuming again the alignment of the molecular dipole moments along the
lines of force of the ion, one can use the equilibrium conditions
(\ref{eq:equ2}) and (\ref{eq:equ3}) to study the profile of dipole
moment and density around the cluster.  Consider first the dipole
moment as the order parameter with an equation of state $W(n,d)$ in
the form of the solid curve shown in Figure 3.  The equilibrium dipole
moment is determined by the condition that the function $W_{\rm
ext}(n,d)$ is a minimum of $d$.  As $W_{\rm
ext}(n,d)=W(n,d)-|q|e^2d/r^2$, the minimum will be located at
different values of $d$ for various values of $r$.  In Fig. 3 the
solid curve gives $W_{ext}(n,d)$ at a distance far from the ion in the
bulk part of the water medium.  The metastable minima $d_1$ and $d_2$
are above the normal ground state $d_0$.  At a distance $r_1$ from the
ion, the polarization term $(-|q|e^2d/r^2)$ proportional to $d$
distorts the function $W_{\rm ext}$ such that the minimum around $d_1$
is on the same level as that around $d_0$, and a state of coexistence
of the two states is possible.  At a shorter distance, the dipole
moment will make a transition to $d_2$. Similarly, at distances closer
than $r_2$, the minima at $d_2$ is pulled down lower than the minimum
around $d_1$, and the system will make a transition to the dipole
moment around $d_2$.  Thus, the dipole moment is $\sim$$d_0$ for $
r>r_1$, is $\sim$$d_1$ for $r_1>r>r_2$ and is $\sim$$d_2$ for $r_2>r$,
and so on.  Thus, in the presence of polymorphic states of water, the
local dipole moment of the cluster abruptly changes as a function of
the radial separation from the ion, corresponding to the different
minima brought down by the polarizing interaction.

One can use a similar argument to discuss the density profile of the
medium around an ion.  The equilibrium density is determined by the
condition that the function $nW_{\rm ext}(n,d)= nW(n,d)-n|q|e^2d/r^2$
is a minimum with respect to a variation in $n$.  If the function
$W_{\rm ext} (n,d)$ possesses secondary minima, then $nW_{\rm
ext}(n,d)$ will also possess multiple minima.  Using arguments similar
to those for the dipole moments, one finds that in the presence of
polymorphic states in density, the local density around a cluster
changes abruptly as a function of the radial distance from the ion.
After equilibrium is reached, water molecules at different distances
will have different densities, being greater at shorter distances from
the ion.

The abrupt changes of density or dipole moment arise only when the
heights of the secondary minima in these two degrees of freedom are
not too large so that they can be pulled down by the polarizing power
of the ion.  If the heights of these secondary minima are very large,
then, for all intents and purposes, the medium is essentially a single
state where the secondary minima play no role.

\section{Cluster Interaction and Fragmentation}

After equilibrium is reached and the clusters formed around the ions,
each cluster remains charged, and there is a Coulombic interaction
between different clusters.  Each cluster has a radius of about 10
\AA~, and the Coulomb interaction energy between two touching clusters
of opposite charge is about 0.7 eV which is still considerable.  It
may provide sufficient attractive interaction for the clusters to
arrange themselves in an orderly manner, forming clusters of greater
sizes (superclusters) with many \xc and \yc entities.  Another
interesting possibility is the formation of a bridge linking the two
clusters of opposite charge between which the dipole moments of the
\h2o molecules line up in the same direction.  It is of interest to
examine in I$_E$ water whether superclusters or bridges of \h2o molecules
are components of the anomalous states in I$_E$ water.  Thus, the study
of these clusters in I$_E$ water will provide information on the
stability and the interactions of the \xc and \yc in aqueous
solutions.

If the liquid is shaken vigorously after equilibrium is reached, then
the cluster will fragment into many small domains.  In each domain, the
dipole moment has been properly aligned by the ion.  If the water
medium has only a single phase, these domains will relax and will
return to the state of the normal water, with the orientation of the
dipole moments of the \h2o molecules becoming randomized again.  No
new anomalous state will be formed as a stable entity.

The situation will be different if there are polymorphic states of
water and the domains which break away from the cluster contain
metastable polymorphic states, which have been produced under the
strong field of the polarizing ion.  Being a polymorphic state stable
under local variations of its order parameters, these domains will not
relax to the normal state of water.  They will remain metastable and
may coalesce with other similar domains.  Fragmented domains of \h2o
can also act as seeds for the growth of greater regions of polymorphic
states.  In this case, anomalous polymorphic states will contain
metastable domains of \h2o molecules for which the dipole moment or
the density may be different from that in the normal state.

\section{Conclusions and Discussions}

The anomalous states of water are peculiar and unexpected.  Their
existence may be connected to the occurrence of metastable polymorphic
states of water at room temperature.  It is therefore important to
confirm or to refute the observations of these anomalous states by
independent experimental investigations.

Looking at the process of the formation of the anomalous states, one
finds that isolated ions have great polarizing power which attracts
water molecules around it.  Stable clusters in free assembly have been
copiously found experimentally and their existence in dilute aqueous
solutions is expected.  The large polarizing power of the ionic charge
leads to great change of the dipole moment and density of the medium
around the ion.  These changes depend on the stiffness of the equation
of state against the variation of density and dipole moments.  If
there are low-lying polymorphic states of the liquid, the polarizing
action of the ion will lead to local densities and local dipole
moments which change abruptly as a function of the separation from
the ion, being greater at shorter distances from the ion.

Fragmentation of the clusters upon vigorous shaking will produce
domains of the liquid where the dipole moments are aligned.  The
coalescence of these domains and the seeding of these regions may
allow the formation of polymorphic states of the liquid if these
polymorphic states are possible metastable configurations of the
liquid.  Pending further experimental confirmations, the anomalous
states of the I$_E$ water may be such a substance.

The presence of the polymorphic state can be examined by looking for
the decay of the metastable I$_E$ state.  It can also be studied by
exciting the I$_E$ state above the barrier which will lead to a change
of the I$_E$ state to the normal state, and will deplete the I$_E$
water population.  One can also measure the dielectric properties of
the I$_E$ water to show an enhanced dipole moment of the I$_E$ water.
The difference in the dielectric constants as discussed by Lo
$et~al.$~\cite{Lo96b} needs to be analyzed theoretically to understand
its implications on the nature of the dipole moments in the anomalous
states.

The foregoing discussions in the last few sections can be carried over
to discuss many other liquids in which the molecules in the liquid
have permanent dipole moments. Because the polarizing interaction for
an aligned dipole depends on $-|q|e^2 d /r^2$, the interaction is
greater, the greater the static electric dipole moment of a single
molecule.  Therefore, the addition of a small amount of ionic compound
on a liquid whose molecules have a large static dipole moment will
lead to a very strong clustering of the molecules around the ion.
They can be well utilized to study polymorphism in these liquids to
see whether domains of anomalous states with properties different from
those in the ground states may exist in these liquids.
\vfill

\pagebreak
\section*{Acknowledgments}
The authors would like to thank Profs. A. A. Chialvo and S. Christian
for helpful discussions.  The research of CYW is sponsored by the
USDOE under Contract DE-AC05-96OR22464 managed by Lockheed Martin
Energy Research Corp.

\end{document}